\documentclass[pra,twocolumn,longbibliography,amsfonts,amssymb,amsmath,floatfix,floats,a4paper]{revtex4-1}

\usepackage [margin=1in]{geometry}
\usepackage{physics}
\usepackage{graphicx}
\begin{document}

\title{Comment on ``Past of a quantum particle revisited''}

\author{Uri Peleg}
\affiliation{Raymond and Beverly Sackler School of Physics and Astronomy, Tel-Aviv University, Tel-Aviv 69978, Israel}
\author{Lev Vaidman}
\affiliation{Raymond and Beverly Sackler School of Physics and Astronomy, Tel-Aviv University, Tel-Aviv 69978, Israel}

\begin{abstract}
The recent criticism of Vaidman's propsal for the analysis of the past of a particle in the nested interferometer is refuted. It is shown that the definition of the past of the particle adopted by Englert et al. [Phys. Rev. A 96, 022126 (2017)] is applicable only to a tiny fraction of photons in the interferometer which indeed exhibit different behaviour. Their proof that all pre- and postselected particles behave this way, i.e. follow a continuous trajectory, does not hold, because it relies on the assumption that it is intended to prove.
\end{abstract}
\maketitle

A recent paper \cite{Berge} (EHDLN) analyzes Vaidman's three-path interferometer with weak path marking \cite{past}  and disagrees with the original claim that the particles have discontinuous trajectories. In Vaidman's narrative the particles propagate along all three paths, which is at odds with the single-path story told by common sense.

In this Comment we  defend Vaidman's narative by showing that EHDLN's argument for a particular single-path story can be repeated equally well for another single path, contradicting (EHDLN's) common sense approach according to which a particle must have a single continuous trajectory.

There is a large body of  literature on this subject \cite{Danan,LiCom,RepLiCom,morepast,Jordan,Sali,SaliCom,Bart,BartCom,Poto,PotoCom,China,ChinaCom,Sok,SokCom,SokComRep,Grif,GrifRep,Nik,NikRep,Hash,HashCom,HashComRep,Dupr,DuprCom,DuprComRep,Zhou,Disapp,Hasegava,NikRR,Eli}. The novel aspect of EHDLN's paper is ``extracting unambiguous which-path information from the faint traces left by an individual particle on its way through the interferometer.''
In the abstract of \cite{Berge} it is stated: ``In our analysis, `the particle's path' has operational meaning as acquired by a path-discriminating measurement.'' This is not the meaning of the past of the particle according to Vaidman's proposal.  Vaidman  argues that there is a consistent way to attribute location to a particle in the past without unambiguous path-discrimination measurements of the faint traces. He {\it defines} that the particle was where these faint traces are present. The information about the traces is obtained either by a calculation, or by a measurement performed on the pre- and postselected ensemble, not by ``unambiguous path-discrimination measurement'' of traces of a single particle.

In fact, if the EHDLN unambiguous measurement is performed and the presence of the trace is detected, Vaidman's approach is in agreement with EHDLN's: the (other) faint traces are present only on EHDLN's continuous path. It might seem then that the disagreement is semantic: a difference in definitions. This can be so, if EHDLN would declare, following Bohr's approach, that the past is defined only in the (very rare) cases when the measurement unambiguously detects the trace of the particle. However,  EHDLN argue that even in the cases when the unambiguous path-discrimination measurement fails to detect the trace, one can still claim, using the ``accounting exercise'' (in section IV.B), that the particle had the common sense single-path trajectory.

The subject of our discussion, the nested interferometer, is a relatively simple setup and the reason why there are so many papers discussing the question: ``Where was the particle?'' is that standard quantum mechanics does not have an answer. For a pre-selected only quantum particle, physicists are reluctant to consider this question, although a possible answer is present: everywhere where the wave function is non vanishing. In the case in which the particle is both pre- and postselected the standard formalism has no answer.

An approach that is sometimes considered as a ``common sense"  interpretation, is Wheeler's proposal \cite{Whee} according to which  we  associate a well defined path with a quantum particle when its wave function spits into wavepackets with well defined trajectories and only one of them connects the source with the detector. A particular tuning of the nested interferometer is an example:  a definite trajectory in Wheeler's sense exists, the path $C$, see Fig. 1. Vaidman argued that the trajectory defined in this way is not helpful, since it lacks operational meaning. He proposed an alternative definition based on an operational meaning: the particle was where it left a (weak) trace. In fact, in most cases, Vaidman's and Wheeler's definitions provide the same picture, but in the case of the nested interferometer, Vaidman's definition yields a weak trace not only in path $C$, but also in the arms of the inner interferometer which are parts of paths $A$ and $B$. Vaidman, together with his collaborators, also performed an experiment \cite{Danan} demonstrating that there is a weak trace of the same strength in paths $C$, $A$ and $B$. EHDLN criticised the experiment since it ``does not make any information available about individual photons ... [and] the data is perfectly consistent with an alternative story: Each photon of a small fraction leaves a discernible trace at checkpoint $A$ or at $B$ or at $C$, while most photons leave no trace at all.''

\begin{figure}
\begin{center}
  \includegraphics[width=8cm]{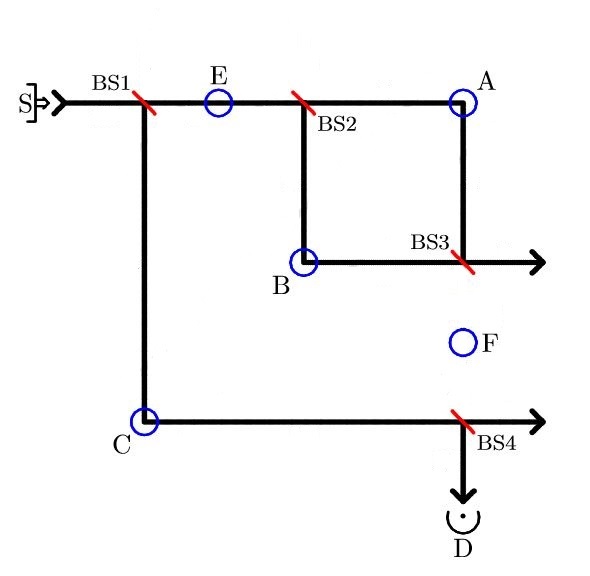}
\end{center}
  \caption{Vaidman's nested interferometer in the EHDLN notation. Inner interferometer is tuned to destructive interference towards $F$.}
  \label{fig:FIG-A}
\end{figure}

Vaidman intentionally avoided measurements providing unambiguous path information of individual photons. Without  measurement interactions with macroscopic amplification, the EHDLN alternative story is impossible in the framework of the standard quantum mechanics. We are in the framework of Sch\"oringer equation which tells us that every photon alters the microscopic state of the environment. Every photon leaves some change (the trace). The EHDLN verification measurement of the presence of this change, the unambiguous path discrimination measurement, can, with some probability, ``erase'' this trace, but without individual measurements, the trace must still be there.

But let us consider the measurements  of individual measuring devices which provide unambiguous path discrimination as EHDLN suggests. In the rare cases when an unambiguous mark is found, EHDLN and Vaidman agree that there is a single continuous trajectory through the detected path. EHDLN would use ``common sense'',
while Vaidman would calculate that, given the detection of the mark, there will be slight changes to the states of the systems of the environment  only on one continuous path, creating a faint trace only on that path.
The controversy arises when the mark is not detected. EHDLN argue that the particle must take the  ``the common sense" path $C$, while Vaidman  claims that the local environment will have a faint trace not only in a continuous path $C$, but also (and of the same strength) in parts of the paths $A$ and $B$ (inside the inner interferometer).

Now we present a simplified version of the EHDLN argument. We  neglect  the terms proportional to $\epsilon$ because the disagreement is not about  differences of order $\epsilon$. Indeed, in figure caption 5   EHDLN claim: ``In the limit $ \epsilon \rightarrow 0$, all particles reach $D$ via $C$.''  Then, the EHDLN argument goes as follows. First is the (correct) observation that introducing an arbitrary phase in path $C$ does not change detection probability in $D$, which is 1/9. From this result, EHDLN concluded that the photons  which reach the detector passing through $C$ are incoherent with those passing through all other paths and therefore, to find the probability in $D$  we have to sum the probability of photons from $C$ with probability of photons from elsewhere. Second, the (correct) calculation of the fraction of photons from the source which pass though $C$ and reach the detector $D$, which is also 1/9. Then, they have argued that the accounting exercise, presented in Section IV~B, shows that there is no room for any other photons and therefore, no photons detected in $D$ were in the inner interferometer.

If the photons passing through $C$ are incoherent with all other photons reaching $D$, the EHDLN argument holds. This incoherence can be arranged, for example,  by adding to Vaidman's setup an unambiguous 100\% efficient  marker on path $C$. We agree that in this case all photons detected at $D$ pass through $C$. The calculation shows that a faint trace due to weak interaction of the photon will appear only on the continuous path passing through $C$. The only error of EHDLN is the claim that the photon's wave packet in $C$ is incoherent with other wave packets reaching $D$. The probability of detection in $D$ is insensitive to the introduction of a phase in $C$ not because of the lack of coherence, but because the other wave packets, passing through $A$ and passing through $B$, interfere destructively.

Probably the most convincing way to show that our simplified version of the EHDLN argument cannot be true is to apply it and reach a contradiction. The EHDLN claim that the photons go solely through $C$ because detection in $D$ is insensitive to the phase change in $C$ and the intensity in $D$ equal exactly to the intensity in an alternative experiment when all paths except $C$ are blocked. However, a simple  calculation shows that the same is true for path $B$. The intensity in $D$ is 1/9 and it  is insensitive to the phase change in $B$ if it is introduced instead of the phase in $C$. And it remains 1/9 if all but path $B$ are blocked. So the EHDLN argument forces us to conclude that the particle took a continuous path solely through $B$ and also, that  it took a continuous path solely through $C$. This contradicts  the ``common sense''.

Our presentation of the  EHDLN argument was missing some details. We omitted all effects of order $\epsilon$, but more importantly, we did not follow the full history of the particle. We only considered the particle leaving the source, being in $C$ or (and)  $B$, and reaching the detector $D$. The EHDLN approach added to the discussion the possibility of a passage through $F$, see their Fig. 5.

In the caption to Fig. 5 it is stated: ``Since that fully accounts for the particles that took the path $BS3 \rightarrow BS4 \rightarrow D$, the inconclusive measurement outcomes (gray) surely identify particles that followed
the path $C \rightarrow BS4 \rightarrow D$.'' So, while we present the distinction between particles reaching $D$ through $C$ and not through $C$, EHDLN consider the distinction between going through $C$ and through $F$, which is the path $BS3 \rightarrow BS4$. They  added a full section (VI) for the analysis of particles entering the inner interferometer and reaching point $F$. This section provided a correct conclusion: every particle reaching point $F$ left an unambiguous mark of its presence in one of the paths, $A$ or $B$. Therefore, a particle passing through  $F$ cannot be one of the particles which left no trace in the markers.

 This supports EHDLN's accounting exercise (Section IV~B): all particles that left no mark had to pass only through path $C$. In this approach there is no contradiction: there is no equivalent proof that all particles detected at $D$ passed through $B$. Indeed,  the setup is not symmetric and there is no  point to be the analogue of $F$.  Trajectories connecting  $A$ with $D$ and $B$ with $D$, pass through $F$, but trajectories going trough $C$ do not pass through $F$. There is no point in the intersection of the trajectories connecting  $C$ with $D$ and $A$ with $D$ which is not on a trajectory connecting $B$ with $D$.

In considering passing through $F$, EHDLN made a tacit assumption: every photon has a continuous trajectory. This assumption contradicts Vaidman's story:  At intermediate times it is present simultaneously in three places $A$, $B$ and $C$, but before and after, it is only in $C$.  Assuming that the photons always  follow a continuous trajectory and adding an unambiguous path discrimination measurement indeed explains the faint traces in $A$, $B$ and $C$ and proves that all particles which left no trace passed through $C$.  But it is a circular argument:  the EHDLN approach``proves'' that Vaidman's story is incorrect by  assuming that it is incorrect.

In summary, Vaidman's analysis of local faint traces in the nested interferometer is correct. His proposal to define the past of the particle as places where it left these faint traces is consistent. EHDLN's definition of the past ``as acquired by a path-discriminating measurement'' provides a different picture: the past is always described by a continuous trajectory. However, the latter applied only to a tiny fraction of the photons passing through the interferometer, those which are identified by unambiguous path-discriminating markers. The proof of EHDLN that all pre- and postselected photons have continuous trajectories and that, therefore, Vaidman's picture is inconsistent, is incorrect. The proof is based on the assumption that all photons reaching the detector have to pass either through $C$ or through $F$. This can be justified only if one assumes that the photons can follow only a continuous trajectory, the statement denied by Vaidman, i.e the statement that has to be proved.

This work has been supported in part by the Israel Science Foundation Grant No. 1311/14 and
the German-Israeli Foundation for Scientific Research and Development Grant No. I-1275-303.14.



\end{document}